\documentclass[twocolumn,english,showpacs,floatfix,amsmath,amssymb,prb,superscriptaddress]{revtex4-1}
\usepackage{graphicx}
\usepackage[colorlinks=true,citecolor=blue,linkcolor=red,linktocpage=true,pagebackref=false]{hyperref}
\usepackage{color,soul}

\newcommand{\be}{\begin{equation}}
\newcommand{\ee}{  \end{equation}}
\newcommand{\ba}{\begin{eqnarray}}
\newcommand{\ea}{  \end{eqnarray}}
\newcommand{\ve}{\varepsilon}
\begin{document}

%\def\openone{\leavevmode\hbox{\small1\kern-4.2pt\normalsize1}}
%%%%%%%%%%%%%%%%%%%%%%%%%%%%%%%%%%%%%%%%%%%%%%%%%%%%%%%%%%%%%%%%%%%%%%%%%%%%%%%
\title{Transport  in quantum spin Hall edges  in contact to a quantum dot }

\author{Bruno Rizzo} 
\affiliation{Departamento de F\'{\i}sica, Facultad de Ciencias Exactas y Naturales, Universidad de Buenos Aires, Pab.\ I, Ciudad Universitaria, 1428 Buenos Aires, Argentina }

\author{Alberto Camjayi} 
\affiliation{Departamento de F\'{\i}sica, Facultad de Ciencias Exactas y Naturales, Universidad de Buenos Aires, Pab.\ I, Ciudad Universitaria, 1428 Buenos Aires, Argentina }

 \author{Liliana Arrachea} 
 \affiliation{International Center for Advanced Studies, UNSAM, Campus Miguelete, 25 de Mayo y Francia, 1650 Buenos Aires, Argentina}

\date{\today}
\begin{abstract}
We study the transport mechanisms taking place in a quantum spin Hall bar with an embedded quantum dot, where electrons localize and experience Coulomb interaction $U$ as well  
as spin-flip processes $\lambda$. We solve the problem with non-equilibrium Green functions. We focus on  the linear response regime and treat the many-body interactions with quantum Monte Carlo.
The effects of $U$ and $\lambda$ are competitive and the induced transport takes place through different channels. The two mechanisms can be switched by changing the occupation of the dot with a gate voltage.
\end{abstract}
\pacs{73.23.-b, 73.63.Kv, 73.21.La}
\maketitle

\section{Introduction}
The existence of metallic states at the boundaries of topological insulators (TI) is one of the most appealing characteristics  of these materials.\cite{Murakami:2004th,Bernevig:2006ij,Hasan:2010ku} In the case of  the two dimensional (2D) systems, the edge states are helical (HES) and exist in the form of Kramers' pairs 
of counter-propagating electron states with opposite spin.\cite{Kane:2005hl,Wu:2006ds,Xu:2006da}
These states are topologically protected \cite{Kane:2005gb} against disorder in the absence of time-reversal symmetry breaking factors such as a magnetic field \cite{Maciejko:2010cl} or magnetic impurities.\cite{Maciejko:2009kw,Tanaka:2011kk,Hattori:2011hj}
As a consequence, the transport is ideally ballistic and each pair of HES supports a 
 perfect conductance quantum $G_0= e^2/h$.
Recent experiments in HgTe quantum wells indeed provided evidence of the existence  of 1D topological HES.\cite{Konig:2007hs,Konig:2008bz,Roth:2009bg,tli}

The  possibility of realizing electron interferometers  in 2D TI  bars, akin to those fabricated with quantum Hall edge states captured a significant attention.\cite{Dolcini:2011dp,Citro:2011fw,romeo,Ferraro:2013,nos,int1,int2,int3,Liu:2011,Ronetti:2016,Probst:2011,Mani:2016}  
Quantum interference  is generated when tunneling processes between different Kramers pairs at opposite sides of the bar take place. 
 The  usual  process is tunneling preserving spin,  but the scenario is much richer when the tunneling with spin-flip also happens. \cite{Dolcini:2011dp,nos} Such scattering process does not break time-reversal symmetry and makes
  an helical interferometer different from two independent copies of a chiral electronic interferometer like those built in the quantum Hall regime.
  
Another relevant feature  is the possibility  of generating  effective back-scattering processes within the same Kramers' pair. This, in turn, generates effective
resistive behavior  with the concomitant departure of the conductance from the ideal quantum limit. A possible mechanism  is the coupling of the HES to magnetic disordered impurities\cite{alt} or quantum dots representing  puddles of the sample.\cite{var,var1}
Indications of such resistive behavior has been actually experimentally observed, which adds motivation to a deeper understanding of this phenomenon. The coupling of helical edge states to quantum dots and magnetic impurities has been addressed in several works.\cite{plat,sch,chei,plee,alt,crep,dolc,syw,var,var1,Posske:2013,Posske:2014,Probst:2015,Rod:2016,Dolcetto:2016} 
A crucial ingredient for a net resistive behavior to take place is an effective anisotropic coupling between the localized spins and the spins of the electrons in the HES. 

%%%%%%%%%%%%%
\begin{figure}
 \includegraphics[clip,width=\linewidth]{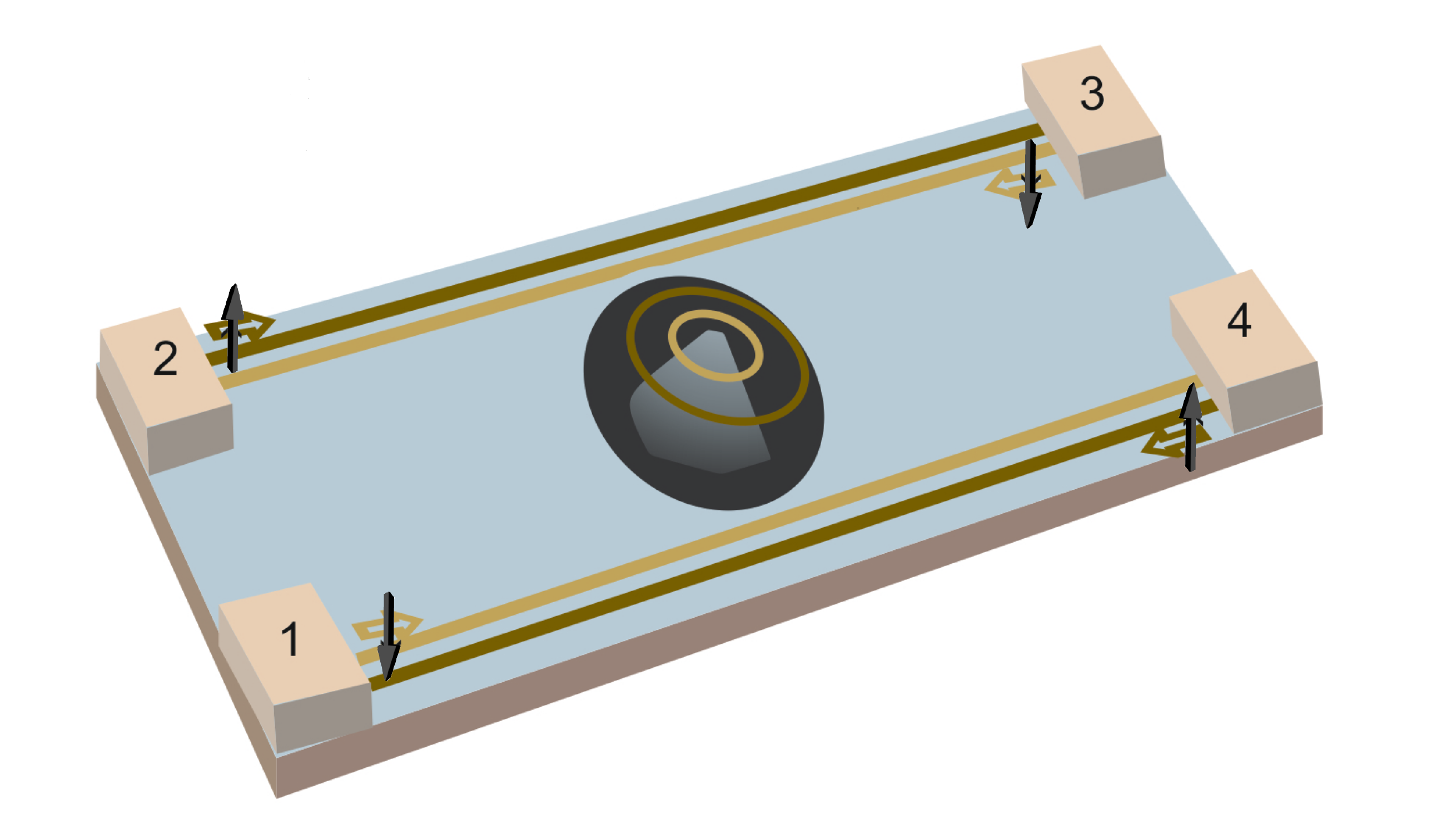}
 \caption{\label{Fig1}(Color online) Sketch for the setup of a quantum dot coupled to the edge channels of a topological insulator. The dot can be generated from the application of a negative gate voltage on the surface of TI. One direct consequence is the formation of localized states around the potential peak that mediates the different edge Kramers pairs. We assume that the coupling between the channels and the dot is punctual.}
\end{figure}
%%%%%%%%%%%%%%

 In the present work we analyze the transport in helical edges of a 2D spin Hall bar with an embedded quantum dot. We focus on 
the combined effect of 
many-body interactions and local spin-flip processes taking place at the dot. The coupling between the edge states with such a  quantum dot gives rise to backscattering and resistance within a Kramers' pair.
It also generates effective
  tunneling
processes between states at opposite edges of the bar preserving and flipping spin.
 We consider the setup sketched in Fig.~\ref{Fig1}, in which a quantum dot in the center of the bar is coupled to the HES. This may represent an antidot generated by a top gate as in quantum Hall systems \cite{ford} or a charge puddle of the sample.   We analyze the impact of the different scattering mechanism that may take place at the dot on the electron transport of this device.
The work is organized as follows. In Section \ref{sec:Model} we introduce the model and in Sections \ref{tran} and \ref{sec:green} the theoretical treatment. Section \ref{sec:results} is devoted to present results. Discussion and conclusions are presented in Section \ref{sec:conclusions}.

\section{Model}
\label{sec:Model}
The full setup of a spin Hall bar with four contacts is sketched in Fig.~\ref{Fig1}. This corresponds to a simplification of a six-terminal setup like the one studied in Ref.~\onlinecite{Roth:2009bg}. The latter setup contains left and right terminals, as well as two at the bottom and two at the top of the bar. Terminals $(1)$ and $(4)$ of the Figure represent the two connected at the bottom, while $(2)$ and $(3)$ represent the two at the top of the bar in the six-terminal configuration.  The bar also  supports a quantum dot in the bulk, which can be generated by locally applying a gate voltage in a small region of the surface
of the TI. As a  consequence of the local voltage, Kramers pairs of edge states are originated enclosing the gate region. When the latter are close to the top and bottom edges of the bar, tunneling processes take place
between the edge states of the dot and those of the bar. The corresponding 
Hamiltonian  reads
 $H=H_{0}+H_{D}+H_{T}$. The Hamiltonian for the electrons in the edge states is:
\be \label{HTI}
H_{0} = \sum_{ \alpha=L, R} \sum_{ \sigma=\uparrow,\downarrow} \int dx  : \Psi_{\alpha,\sigma}^{\dagger}(x) {\cal D}_{\alpha, x} \Psi_{\alpha ,\sigma}(x) : ,
\ee
where $x$ denotes a longitudinal coordinate and the integral is over the length of the edges. $\Psi_{\alpha,\sigma}(x)$ are fermionic fields describing electrons moving right and left  along the edge 
($\alpha=R,L$, respectively)  with spin $\sigma= \uparrow, \downarrow$. 
Here ${\cal D}_{\alpha, x}= \mp i \hbar v_F  \partial_x$ for $\alpha=L,R$ respectively while $:\!O\!:$ denotes normal ordering with respect to the state where all the states are filled up to the Fermi energy and 
 $v_F$ is the Fermi velocity. The quantum dot  is modeled by an Anderson impurity model, where the helical edge states circulating around the local gate are represented by a localized electron level 
 with the same energy $\varepsilon_0$ for electrons with up and down spin component,
 \be \label{Hd}
H_{D}=\sum_{\sigma=\uparrow,\downarrow} \left[ \varepsilon_0 d_{\sigma}^{\dagger}d_{\sigma}+ \frac{U}{2} n_{\sigma}n_{\overline{\sigma}} +\lambda d_{\sigma}^{\dagger}d_{\overline\sigma}\right],
\ee
where $U$ is the Coulomb repulsion for the electrons in the quantum dot and $\overline{\sigma}$ denotes the spin opposite to $\sigma$. The value of $\varepsilon_0$ is controlled by the local gate voltage. We also include a local spin flip 
process in the dot by a phenomenological parameter $\lambda$.\cite{flip1,flip2,flip3,flip4,flip5}
Such a process may mimic  the interaction of the localized spins at the dot with nuclear spins of
the sample.\cite{plat} Another mechanism to realize such a term is by directly applying an external magnetic field transverse to the direction of the spin-orbit interaction of the 2D topological insulator. Finally, the term 
\ba \label{HT}
H_{T} & = & \sum_{\alpha=R,L} \sum_{\sigma=\uparrow,\downarrow} \int dx \, \left[ \Gamma(x) \Psi_{\alpha,\sigma}^{\dagger}(x)d_{\sigma}  +  h. c. \right],
\ea
with $\Gamma(x)= \frac{v_F\hbar}{\sqrt{ d}}\gamma \delta(x-x_{0})$ represents the spin-preserving tunneling between the dot and the edge states. Here $ d$ is a characteristic length of the contact.

\section{Transport processes of an edge-state terminal}
\label{tran}
We consider the configuration indicated in Fig.~\ref{Fig1} where each of the  four corners of the bar is contacted to independent reservoirs at which separate bias voltages can be applied. These four voltages enable an independent control on the injection of electrons into the four edge states. In what follows, we identify the terminal contacting the reservoir $V_l$ with the
label $l$. Each of these terminals hosts a Kramers pair of edge states containing a state incoming the contact ($l+$) and another one outgoing it ($l-$). 
Hence, the current flowing through a given terminal $l$ is defined by the difference between the incoming and outgoing density of electrons \cite{nos}
\be \label{cur}
I^l= 
-i e v_F \left[G^<_{l+,l+}(x_l,x_l;t,t) - G^<_{l-,l-}(x_l,x_l;t,t)\right].
\ee
There is a one to one identification between the labels $l{+(-)}$ and the labels the helicity ($\alpha=R,L;\sigma=\uparrow,\downarrow$). For instance, if we focus on
the terminal $3$, we see that $3+ \equiv R, \uparrow$ and $3- \equiv L, \downarrow$.

We define the Green function
$G^{<}_{\alpha \sigma, \alpha^{\prime} \sigma^{\prime}}(x, x^{\prime};t,t^{\prime})=
i \langle \Psi_{\alpha^{\prime},\sigma^{\prime}}^{\dagger}(x^{\prime},t^{\prime}) \Psi_{\alpha,\sigma}(x,t) \rangle$
and we present details on the calculation  of this function in
  Appendix \ref{sec:appendix}. 
  The resulting
expression of the current can in general be decomposed as follows
\be \label{current2}
I^{l}= I_0^l - I^l_{sp} - I^l_{sf} - I^l_{b}+I^l_{ne}.
\ee
The first term, $I_0^l=\frac{e}{\hbar}  \int_{-\infty}^{+\infty} \frac{d\omega}{2 \pi}  \left[f_{\alpha\sigma}(\omega)-f_{\overline\alpha\overline\sigma}(\omega) \right]$ corresponds to  purely ballistic transport through the terminal $l$ in the absence of any coupling to the quantum dot. The associated conductance is $G_0=e^2/h$. The other terms are due to the coupling to the quantum dot and tend to decrease the conductance with respect to the quantum limit $G_0$. The contributions  $I^l_{sp}$ and  $I^l_{sf}$ are due to the tunneling between different Kramers' pairs through the quantum dot preserving and flipping the spin respectively. The component $ I^{l}_{b} $ is an intra-pair backscattering leading to an effective resistance in the terminal. The last term,  $I^l_{ne}$ is a non-equilibrium contribution with components on  the previous three channels. 
The explicit expressions for these different contributions are
\ba\label{corriente1}
 I^l_{sp} &=& \frac{e}{h} \int_{-\infty}^{+\infty}d\omega\; \mathcal{T}_{\sigma,\sigma}^l(\omega) 
\left[f_{\alpha\sigma}(\omega)-f_{\overline\alpha\sigma}(\omega) \right] ,  \nonumber \\
I^l_{sf} &= &\frac{e}{h} \int_{-\infty}^{+\infty}d\omega\;  \mathcal{T}_{\sigma,\overline\sigma}^l(\omega)
\left[f_{\alpha\sigma}(\omega)-f_{\alpha\overline\sigma}(\omega)\right], \nonumber \\
I^l_{b}  &= &\frac{e}{h} \int_{-\infty}^{+\infty}d\omega\; \mathcal{T}_{\sigma,\overline\sigma}^l(\omega)
\left[f_{\alpha\sigma}(\omega)-f_{\overline{\alpha}\overline\sigma}(\omega) \right] ,\nonumber \\
 I^l_{ne} & = & \frac{e}{h} \sum_{\sigma^{\prime}=\sigma,\overline{\sigma}} \int_{-\infty}^{+\infty}d\omega\;
\frac{d}{\gamma^2 v_F  \hbar} \mathcal{T}_{\sigma, \sigma^{\prime}}^l(\omega) \Lambda_{\sigma^{\prime}}(\omega).
\ea
The functions 
\be \label{tes}
\mathcal{T}_{\sigma,\sigma^{\prime}}^l(\omega)=\left(\frac{v_F \hbar}{\sqrt{d}} \gamma\right)^{4} |g_{\alpha\sigma}(x_l,x_0;\omega)|^{2}|G_{\sigma,\sigma^{\prime}}^{R}(\omega)|^{2},
\ee
characterize scattering processes where the electrons tunnel from the edge $\alpha \sigma$ to the quantum dot, eventually flip the spin (for $\sigma^{\prime}= \overline{\sigma}$) and tunnel again to the edge
$\alpha^{\prime} \sigma^{\prime}$.  The Fermi function 
$f_{\alpha\sigma}(\omega)$  depends on the chemical potential and the temperature of the reservoir injecting the electrons into the edge $\alpha \sigma$. The 
function $g_{\alpha\sigma}^{}(x_l,x_0;\omega)$ is the retarded Green function for the free edge state (see Eq.~\ref{g0r1}) while $G_{\sigma,\sigma^{\prime}}^{R}(\omega)$ is the retarded Green function of the quantum dot coupled to the edge states. In the most general case, this function corresponds to the fully interacting quantum dot out of equilibrium. The
function
\be
 \Lambda_{\sigma}(\omega)=   2 f_{\alpha \sigma}(\omega) \mathcal{I}m\{\Sigma_{\sigma}^{R}(\omega)\}+i\Sigma_{\sigma}^{<}(\omega)
 \ee
is defined by the retarded and lesser components of the self-energy due to the Coulomb interaction at the dot, respectively, $\Sigma_{\sigma}^{R}(\omega)$ and $\Sigma_{\sigma}^{<}(\omega)$.  This function vanishes close to equilibrium, where all the chemical potentials and temperatures of the reservoirs are the same. In fact,  the following fluctuation-dissipation relation holds:
$2f (\omega)\mathcal{I}m\{\Sigma_{\sigma}^{R}(\omega)\}=-i\Sigma_{\sigma}^{<}(\omega)$. Hence, this process contributes to the current with a leading order $\propto V^2$, being $eV$ 
the bias voltage applied at the terminals with respect to the reference chemical potential $\mu$. Instead, 
the other three terms  $I^l_{sp}$, $I^l_{sf}$ and $ I^{l}_{b} $ contribute at the linear order in the applied voltage $V$.  In what follows, we will focus on small bias voltages where these contributions dominate. We will, thus, neglect the effect of $I^l_{ne}$ and the Green function $G_{\sigma,\sigma^{\prime}}^{R}(\omega)$  will be evaluated in the equilibrium system, as explained in Section \ref{sec:green}.

For low temperatures and small bias voltages, the transport properties of the edge states  in this setup are completed defined by the behavior of the function $\mathcal{T}_{\sigma, \sigma^{\prime}}^l(\omega)$ close to the
reference chemical potential $\mu$. Because of the symmetry of this problem, the diagonal matrix elements of this matrix are identical, $\mathcal{T}_{\uparrow, \uparrow}(\omega)=\mathcal{T}_{\downarrow, \downarrow}(\omega)$ and
also $\mathcal{T}_{\uparrow, \downarrow}(\omega)=\mathcal{T}_{\downarrow, \uparrow}(\omega)$. From Eq.~(\ref{g0}) we directly see that the latter matrix elements vanish for $\lambda=0$. Hence,  if  we focus on a particular terminal 
where the incoming edge is $\alpha, \sigma$ (for instance the terminal $l=3$  at the upper right corner, which has $\alpha=R$ and $\sigma=\uparrow$), the effect of the spin-flip process will contribute to decrease the  corresponding conductance due to backscattering processes when there is a bias applied from left to right ($\mu_1=\mu_2=eV+\mu$ and $\mu_3=\mu_4=\mu$).
A similar behavior will be observed for the case of  a bias applied from bottom to top ($\mu_1=\mu_3=\mu$ and $\mu_2=\mu_4=eV+\mu$) due to the inter-edge processes. On the other hand, in a configuration of voltages
with $\mu_1=\mu_2=\mu_4=\mu+eV$ and $\mu_3=\mu$, we would get a current in the $l=3$ terminal only due to the transmission without spin flip. In this way, a suitable selection of the voltages applied at the terminals will enable to gathering information of the different types of transport processes through the quantum dot. The transport behavior  is fully described by the behavior of the two independent components of the transmission function $\mathcal{T}_{||}= \mathcal{T}_{\sigma,\sigma}(\mu)$ and $\mathcal{T}_{\perp}= \mathcal{T}_{\sigma,\overline{\sigma}}(\mu)$. 

\section{Green function of the quantum dot}
\label{sec:green}

In the previous section we have shown that the transport properties can be completely characterized within linear response in terms of two transmission functions, which depend of the
retarded Green functions  of the isolated edges and of the fully interacting quantum dot coupled to the edge states in equilibrium. The first ones can be analytically calculated and the explicit expressions are given in
Appendix B. In this section we explain how to evaluate $G_{\sigma,\sigma^{\prime}}^{R}(\omega)$. For the case of a non-interacting quantum dot, this function can be analytically evaluated while
for the interacting case, we rely on numerical quantum Monte Carlo simulations. Below, we consider the two cases  separately.
\subsection{Quantum dot without Coulomb interaction}
In the case where $U=0$ it is possible to calculate the retarded Green function of the quantum dot coupled to the edge states analytically. In fact, we can readily solve the
Dyson equation for the retarded Green function in this limit, by  defining the matrix $\hat{G}^0(\omega)$ with matrix elements $G^0_{\sigma,\sigma^{\prime}}(\omega) \equiv G^R_{\sigma,\sigma^{\prime}}(\omega)$.
The inverse of that matrix is 
\be \label{g0}
[\hat{G}^0(\omega)]^{-1}=\left( \omega-\varepsilon_0 - \Sigma_0(\omega) \right) \hat{\sigma}_0 -  \lambda  \hat{\sigma}_x.
\ee
where $\hat{\sigma}_0$ is the $2\times 2$ unit matrix and $\hat{\sigma}_{x}$ is the $x$-Pauli matrix.
We have also introduced the hybridization self-energy
\be \label{sigma0}
 \Sigma_0(\omega) = \left(\frac{\hbar v_F}{\sqrt{d}} \gamma\right)^2 \sum_{\alpha=L,R} g_{\alpha}(\omega) =- i \Gamma^0 ,
\ee
where $g_{\alpha}(\omega) \equiv g_{\alpha\sigma}(x_0,x_0;\omega) $ is the Green function of the free helical edge (the spin label was omitted since it is the same for both spin components) and $\Gamma^0= \gamma^2\hbar v_F/d$.

\subsection{Quantum dot with Coulomb interaction}
For the fully interacting case, $U \neq 0$, we calculate the Green function in the Matsubara representation by recourse to quantum Monte Carlo. To this end we define the matrix 
$\hat{G}(i \omega_n)$ which is the inverse of
\be \label{matsu}
[\hat{G}(i \omega_n)]^{-1}=[\hat{G}^0(i \omega_n)]^{-1} -  \hat{\Sigma} (i \omega_n),
\ee
where $\hat{G}^0(i \omega_n)$ is the Green function of the non-interacting quantum dot coupled to the leads (\ref{g0}) in the Matsubara axis while
$\hat{\Sigma} (i \omega_n)$ is the self-energy matrix due to the Coulomb interaction $U$. These functions are calculated with quantum Monte Carlo by using the ``Continuous-time method'' of Refs.~\onlinecite{mc1,mc2,mc3,mc4}.
 In this method the Anderson impurity coupled to an arbitrary bath of free fermions can be computed in the Matsubara axis at finite temperature. The bath, which in our case
 is represented by the function (\ref{sigma0}) is an input of this algorithm.
The retarded functions and the corresponding transport properties can be calculated by  performing analytic continuation to the real frequency axis of Eq.~(\ref{matsu}).\cite{mc5}
Here we focus on the features close to $\omega=\mu$, which are the relevant ones to evaluate the transmission functions $\mathcal{T}_{||}$ and $\mathcal{T}_{\perp}$ and we 
use the methodology of Refs.~\onlinecite{mc6,mc7}.

\section{Results}
\label{sec:results}
As discussed in Section~\ref{tran} the transport properties of the setup are defined by the behavior of the components of the transmission function preserving and flipping spin,
respectively,  $\mathcal{T}_{||}$ and $\mathcal{T}_{\perp}$. In this section we analyze the effect of the different parameters of the model on the behavior of these functions. The relevant parameters are 
$\lambda$, $U$ and $\varepsilon_0$. 

\begin{figure}[ht]
 \includegraphics[clip,width=\linewidth]{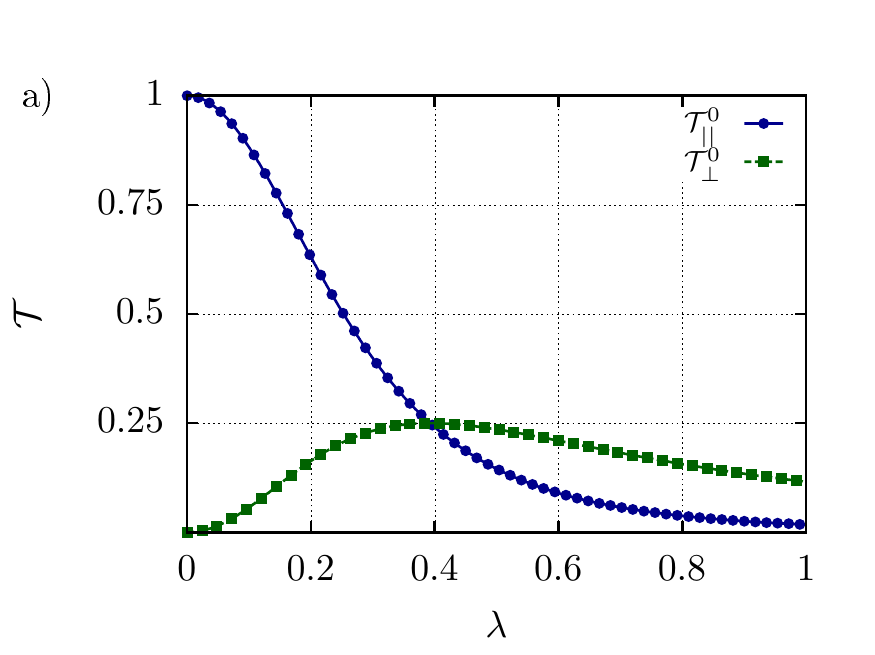}\\
 \includegraphics[clip,width=\linewidth]{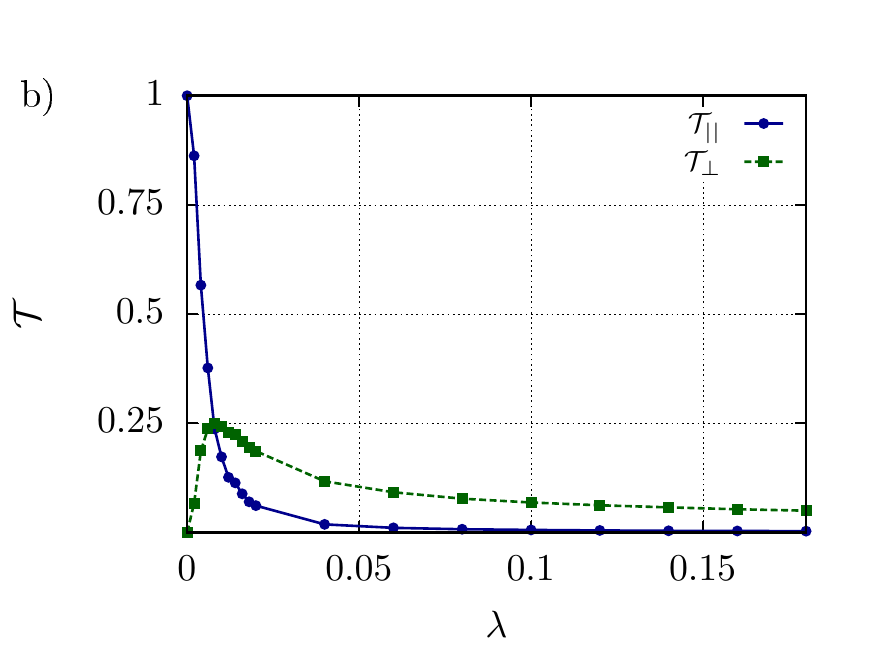}
 \caption{\label{Fig2}
Transmission functions at Fermi level vs the spin-flipping fluctuations $\lambda$ for:  a) Non-interacting model ($U=0$) with $\ve_0=0$ and $\Gamma^0=\pi/8$ and b) Interacting model with $U=4$, $\ve_0=-U/2$ and $\Gamma^0=\pi/8$. While (a) does not depend on temperature, the temperature  in the case (b)  is  $T=0.0025$,  being the Kondo temperature is $T_K \simeq 0.013$ for these values of $\Gamma^0$ and $U$. All energies are expressed in units of $v_F \hbar/d$.
}
\end{figure}

We fix the mean chemical potential of the reservoirs to $\mu=0$ and start by fixing the local gate voltage  of the dot to  $\varepsilon_0=-U/2$. 
In the non-interacting case ($U=0$), this corresponds to the resonant level aligned with the Fermi energy of the edge states. In the interacting case, this corresponds to the dot occupied with a single electron, which is the typical scenario for the Kondo effect to take place. This effect is characterized by the coupling of the spin of the electron localized at the quantum dot with spins of the electrons of the edge states to form a singlet. This is accompanied by a fluctuation in the occupation, which effectively results in an electron resonance and  a  perfect transmission through the dot in the spin-preserving channel. This mechanism takes place only at low temperatures, below the so called Kondo temperature.\cite{kondo,kondoexp} 
In what follows, we fix the effective tunneling rate and focus on $U=4$ and
$\Gamma^0 = \pi/8$, which corresponds to a Kondo temperature $T_K\simeq 0.013$. The values of these quantities are expressed in units of $v_F \hbar/d$.
In Fig.~\ref{Fig2} we show the behavior of the transmissions as the spin-flip amplitude $\lambda$ is varied. The top panel corresponds to the non-interacting quantum dot and the bottom panel to the quantum dot in the Kondo regime. The same global behavior is observed in both cases. Namely,
a decrease of the transmission preserving spin as the transmission with spin flip increases. In the non-interacting case, it is possible to write an analytical expression of the transmissions. In fact the different 
matrix elements of Eq.~(\ref{g0}) read
\ba
G^{0}_{\sigma,\sigma}(\omega) & = & \frac{1}{g^{-1}(\omega) - \lambda^2 g(\omega)},\nonumber \\
G^{0}_{\sigma, \overline{\sigma}}(\omega) & = & \frac{\lambda}{(g^{-1}(\omega))^2 - \lambda^2},
\ea
with $g^{-1}(\omega)= \omega - \varepsilon_0 + i \Gamma^0 $. Substituting these expressions in (\ref{tes}) it is found that  $\mathcal{T}_{||}$ is a decreasing function of $\lambda$ with
maximum $\mathcal{T}_{||}=1$ for $\lambda=0$ where $\mathcal{T}_{\perp}=0$. Instead, $\mathcal{T}_{\perp}$ has a maximum at $\lambda= \Gamma^0$, where $\mathcal{T}_{\perp}= 1/4$. For the interacting case within the Kondo regime, it is natural to find a similar behavior, since this is characterized by a resonance, as in the non-interacting case. This is indeed what we observe in Fig.~\ref{Fig2}b. The results shown in this figure correspond to a temperature much lower than the Kondo temperature and
basically correspond to the $T=0$ limit. In fact notice that for $\lambda=0$ the function $\mathcal{T}_{||}$ achives exactly the unitary limit.
 The scale at which the maximum of $\mathcal{T}_{\perp}$ and the strong decrease of $\mathcal{T}_{||}$ take place is, however, not set by the hybridization width $\Gamma^0$ as in the non-interacting case, but by the Kondo temperature $T_K$. This is because the width of the Kondo resonance 
is $\propto T_K$ and because the spin-flip processes and the formation of the Kondo singlet are competitive effects. Hence, for $\lambda \sim k_B T_K$ ($k_B$ is the Boltzmann constant), the spin-flip process become dominant. 

\begin{figure}[ht]
 \includegraphics[clip,width=\linewidth]{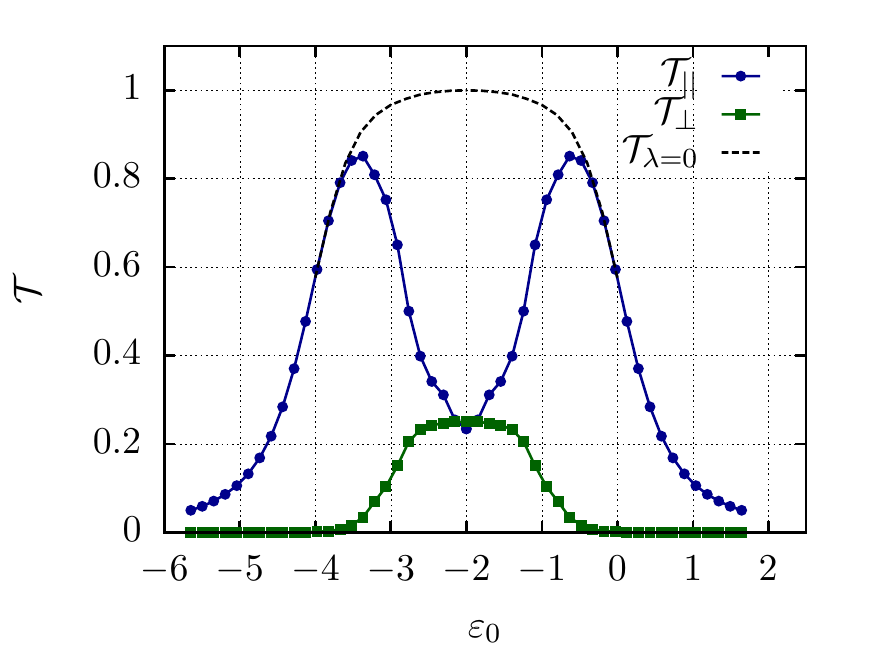}
 \caption{\label{Fig3}
Spin-preserving transmission $\mathcal{T}_{||}$ and spin-flipping transmission $\mathcal{T}_{\perp}$ at the Fermi energy vs the dot level $\ve_{0}$. We fixed a low temperature $T=0.0025$ to tune in the Kondo regime, setting $\lambda=0.008$, $U=4$ and $\Gamma^0=\pi/8$.      
}
\end{figure}

The effect of the gate voltage in the Kondo regime with spin-flip processes is analyzed in Fig.~\ref{Fig3} for a particular value of $\lambda$ close to the one for which $\mathcal{T}_{\perp}$ achieves the maximum  for
the symmetric configuration ($\varepsilon_0 \sim -U/2$)
and a temperature much lower than the Kondo temperature. Interestingly, for values of $\varepsilon_0 \sim -U \mbox{ and }0$, we observe a strong decrease of  $\mathcal{T}_{\perp}$ while $\mathcal{T}_{||}$ displays two maxima 
approaching the unitary limit. The case without spin flip ($\lambda=0$) is shown in dashed lines for comparison. It is characterized by the plateau with unitary transmission within the full interval 
$-U \leq \varepsilon_0 \leq 0$. These features can be understood by noticing that the gate voltage shifts the level of the quantum dot with respect to the Fermi energy of the edge states, leading to a change in
its occupation. For $\varepsilon_0 =-U/2$ the dot is singly occupied, which is the optimal configuration for the Kondo effect to develop. As $\varepsilon_0$ departs from this value, the dot tends to be empty as
$\varepsilon_0 \rightarrow 0$ or doubly 
occupied as $\varepsilon_0 \rightarrow -U$. On the other hand, the spin-flip term is effective only when the dot is singly occupied, in which case it competes with the Kondo effect. In this way, as the gate voltage
approaches the limiting values $\varepsilon_0=0 \mbox{ and } -U$, the spin-flip processes become ineffective and the transport takes place through the two levels separated by the energy $U$ of the Coulomb-blockade regime. This is reflected in the behavior of the transmission functions, by the decreasing amplitude of $\mathcal{T}_{\perp}$ as $\varepsilon_0$ departs
from $-U/2$ and the increment of $\mathcal{T}_{||}$ to the limit of perfect transmission at  $\varepsilon_0=0, -U$ observed in Fig.~\ref{Fig3}.

\begin{figure}[ht]
\includegraphics[clip,width=\linewidth]{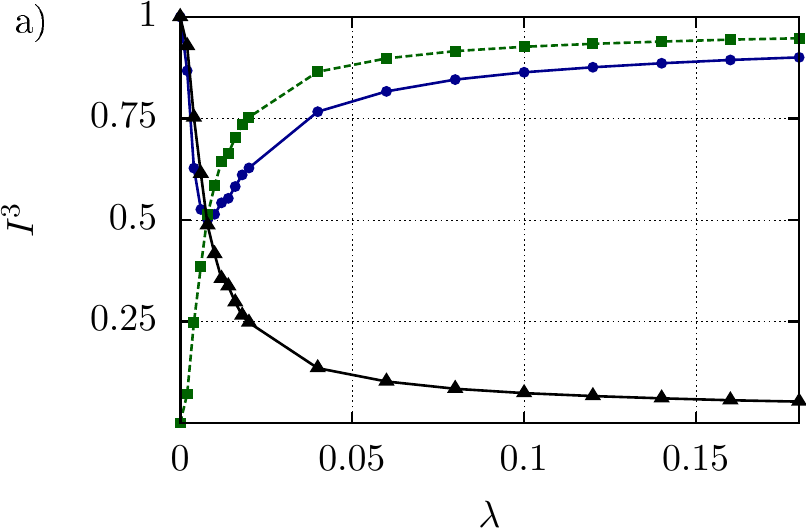}\\
\includegraphics[clip,width=\linewidth]{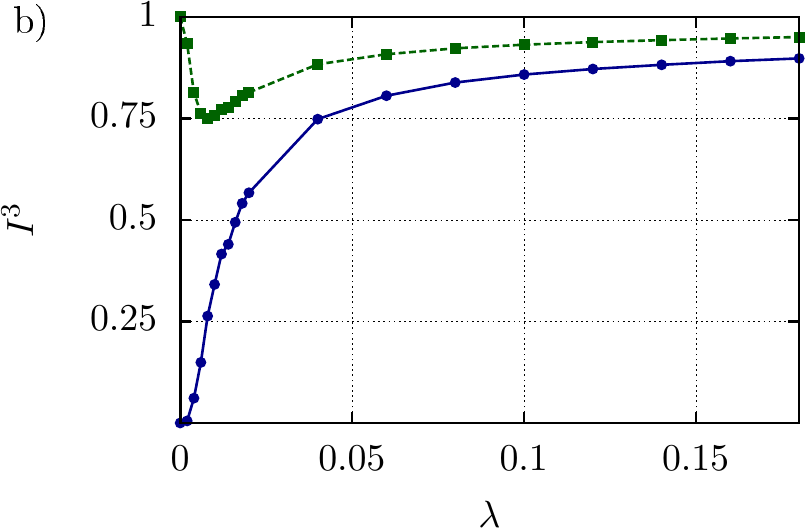}
 \caption{\label{Fig4}Current in units of $e^2V/h$ measured in terminal $3$ vs the spin flipping parameter. a) Bias voltage configuration: $\mu_2=\mu_4=eV$ and $\mu_1=\mu_3=0$ (blue straight line with circles), $\mu_1=\mu_2=eV$ and $\mu_3=\mu_4=0$ (green dashed line with squares), $\mu_1=\mu_4=eV$ and $\mu_2=\mu_3=0$ (black solid line with triangles) b) $\mu_2=eV$ and $\mu_1=\mu_3=\mu_4=0$ (blue solid line with circles), $\mu_1=\mu_2=\mu_4=eV$ and $\mu_3=0$ (green dashed line with squares).
The rest of the parameters are the same as in Fig.~\ref{Fig3}}
\end{figure}

We analyze in Fig.~\ref{Fig4} the behavior of the current measured at a given terminal as a function of the amplitude of the spin-flip processes. We consider the quantum dot in the half-filled configuration corresponding to $\varepsilon_0=-U/2$ and we focus on the terminal at the top right corner of the 2D spin-Hall bar, which is labeled with the index $l=3$ (see Fig.~\ref{Fig1}). We consider different voltage configurations  in order to separate and/or combine the different contributions of the transmission processes through the quantum dot. We begin by considering a bias configuration inducing colliding processes at the quantum dot. This corresponds to $\mu_2 = \mu_4 =eV$ and $\mu_1 = \mu_3 = 0$. The corresponding current is shown in the plot with blue circles of Fig.~\ref{Fig4}a). 
For  $\lambda=0$ the current achieves its ballistic limit $I_0= e^2 V/h$. This is because, in the absence of spin-flip processes at the dot, 
 the particle flow between  HES with same spin projection is forbidden due to the Pauli exclusion principle. In fact, as in this case $\mathcal{T}_{\perp}=0$, the only component that could eventually contribute is 
 the spin-preserving one $I^l_{sp}$, which depends of $\mathcal{T}_{||}$. However, we see
 from Eq.~(\ref{corriente1}) that this component also depends on  the difference of Fermi functions which  vanishes for this voltage configuration. 
With increasing spin fluctuations at the quantum dot, the transmission with spin-flip  $\mathcal{T}_{\perp}$  becomes active, opening the conducting channels $I^l_{b}$ and $I^l_{sf}$. Hence, as $\lambda$ increases,  the net current decreases with respect to
the ideal quantum limit $I_0$, achieving a minimum for the 
 value of $\lambda$ at which $\mathcal{T}_{\perp}$ has a maximum. As $\lambda$  increases further, the spin fluctuations increase and deteriorate the low energy resonance between the dot and HES. This is reflected in a smaller amplitude of $\mathcal{T}_{\perp}$, which leads to and increment of the current towards the quantum limit $I_0$. 
For  $\mu_1=\mu_2=eV$ and $\mu_3=\mu_4=0$ the bar is biased from the left to the right. The corresponding behavior of the current in the terminal $l=3$ as a function of $\lambda$ is shown in  green dashed plot with squares of Fig.~\ref{Fig4}a). For $\lambda=0$ the spin preserving current through the quantum dot  is equal to the quantum limit $I^l_{sp}=I_0$ due to the perfect transmission $\mathcal{T}_{||}=1$. This contribution, exactly cancels the current injected by the terminal $2$ leading to a net vanishing current $I^3$. 
As $\lambda$ increases, $\mathcal{T}_{||}$ decreases abruptly and new  channels open with finite $\mathcal{T}_{\perp}$. The combination of the two contributions, however,  is not enough to cancel the ballistic current and a finite net current flows through the terminal $3$, which  increases in magnitude for increasing $\lambda$. Finally, the plot in black lines with triangles of Fig.~\ref{Fig4}a shows the behavior of the current  $I^3$ for a bias applied from bottom to top, which corresponds to considering 
 $\mu_1=\mu_4=eV$ and $\mu_2=\mu_3=0$. In this case, the  maximum current is found for  $\lambda=0$ due to perfect spin preserving transmission from the lower to upper channel. At finite values of $\lambda$, the contributions $I^l_{b}$ and $I^l_{sf}$ play a role, leading to a decreasing net current. 
 
 A similar analysis can be done for other voltage configurations corresponding to biasing one of the terminals against the other three. Examples are shown in 
  Fig.~\ref{Fig4}b). The case where   $\mu_2=eV$ and $\mu_1=\mu_3=\mu_4=0$ is shown in the blue-line plot with circles. For $\lambda=0$ the perfect transmission $\mathcal{T}_{||}$ 
  through the quantum dot  causes a vanishing flow towards the terminal $3$. The effect of $\lambda$ is to decrease $\mathcal{T}_{||}$, resulting in an increasing net current $I^3$. The other 
  configuration shown in green lines with squares in the Fig.~\ref{Fig4}b) corresponds to  $\mu_1=\mu_2=\mu_4=eV$ and $\mu_3=0$. This voltage configuration is particularly  interesting because 
  it allows for measuring the effect of the backscattering component $I_b$ independently from $I_{sf}$. In fact $I_b$ vanishes for $\lambda=0$ corresponding to the maximum 
  $I^3=I_0$, while the departure of $I^3$ from the perfect ballistic limit is precisely the backscattering component $I_b=I_0-I^3$.

\begin{figure}[ht]
 \includegraphics[clip,width=\linewidth]{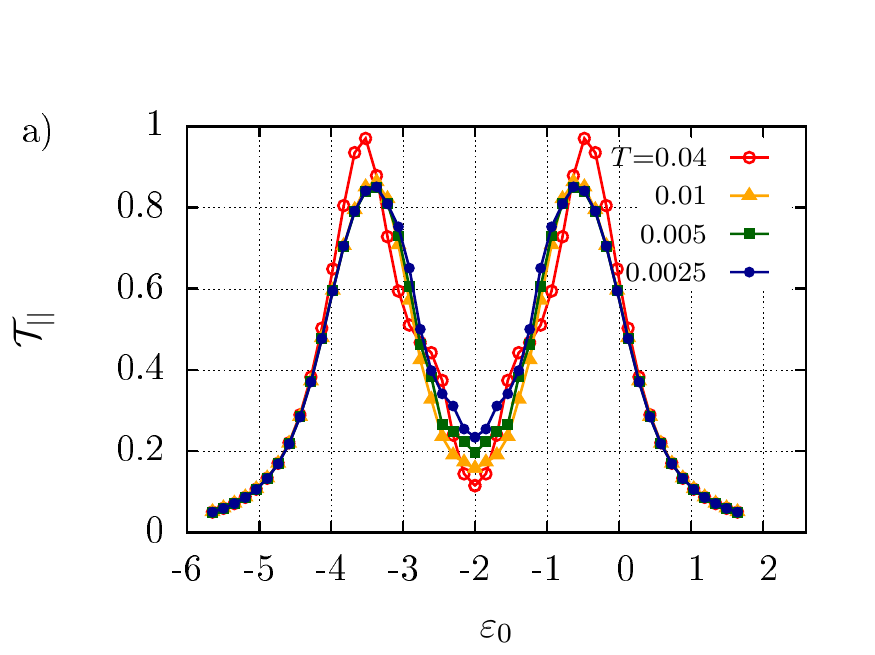}\\
 \includegraphics[clip,width=\linewidth]{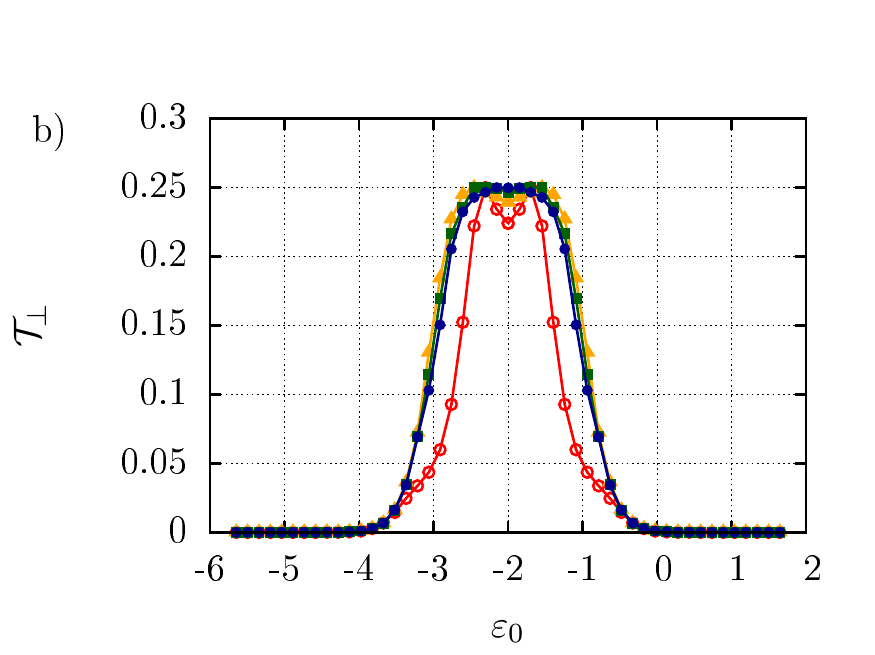}
 \caption{\label{Fig5}
Transmissions functions at the Fermi energy vs the dot level $\ve_{0}$ for different temperature regimes. a) Spin-preserving transmission  $\mathcal{T}_{||}$. b) Spin-flipping transmission $\mathcal{T}_{\perp}$ for the same parameters of Fig~\ref{Fig3}.}
\end{figure}

We close this section by analyzing the effect of the temperature in the features described in Fig.~\ref{Fig3}. This is shown in Fig.~\ref{Fig5} within a range of temperatures below and above the Kondo temperature $T_K$. We see that the effect of the temperature is similar to the effect of the spin flip processes regarding the behavior of the spin-preserving transmission 
$\mathcal{T}_{||}$, which tends to become smaller close  $\ve_{0} =-U/2$ as $T$ increases. Interestingly , the structure of $\mathcal{T}_{\perp}$ is more robust against changes in temperature. This indicates that the terms involving current transmission with spin inversion $ I_{sf} $ and the effective resistance $I_{b}$ depends weakly on the temperature within  the wide range of temperatures explored.
% \textcolor{red}{For $T>T_K$, $\mathcal{T}_{\perp}$ becomes narrower while the maximum tends to split into two peaks}.

\section{Conclusions} 
\label{sec:conclusions}
We have analyzed the transport properties of the edge states of a 2D spin-Hall bar in tunneling contact to a quantum dot where electrons are confined and experience Coulomb interaction $U$ as well as local
spin flip processes $\lambda$. The occupation of the dot can be changed by means to a locally applied gate voltage $\varepsilon_0$. In the singly occupied quantum dot, under the Kondo temperature, the Kondo effect takes place along with the spin-flip processes. These two mechanisms are competitive and contribute to 
the transport along different channels.

The Kondo effect contributes to transport between two different Kramers pairs through the quantum dot without flipping the spin. The spin-flip term contributes to the effective tunneling with spin flip between the two Kramers pairs as well as within the same Kramers pair. The latter corresponds  to an effective inter-pair backscattering and resistive behavior.
By changing the gate voltage it is possible to change the occupation of the quantum dot. Away from singly occupancy, the spin-flip term becomes ineffective and close to $\epsilon_0=-U \mbox{ and }0$, the transport takes place in the Coulomb blockade regime  through the spin-preserving channel. The Kondo and the spin-flip processes are competitive, and the effect of one dominating over the other can be manipulated by the occupation. Hence, the gate  voltage plays the role of a switch to select the spin-preserving or the spin-flip tunneling processes. Such mechanism can be used to design helical interferometers like the ones discussed in 
Refs.~\onlinecite{Dolcini:2011dp,Citro:2011fw,romeo,Ferraro:2013,nos,int1,int2,int3}. In most of these references, the non-trivial effects on the conductance behavior were a consequence of 
tunneling with spin-flip between the HES. In our case, we focussed on local spin-flip processes at the quantum dot and concluded that this provides a channel for an effective tunneling
process with spin-flip between the HES.  In the case of coexisting both types of mechanisms, the combined effects with the Coulomb interaction as well as the consequences on the transport behavior would be qualitatively the same as the ones discussed in the previous section.

If we consider that typical realistic values for the Coulomb interaction and the hybridization function are similar to those in semiconductors, \cite{kondoexp}
$U \sim 1 - 1.3 \mbox{ meV}$ and $\Gamma \sim 0.1 \mbox{ meV}$, then a very small value of the spin-flip parameter $\lambda$ would lead to dramatic consequences in the transport properties below the Kondo temperature.
For instance, notice that  the plots of Fig.~\ref{Fig3} correspond to $\lambda \sim 2 \times 10^{-3} U \sim 2 \,\mu \mbox{eV}$. If, instead of an antidot as in the sketch of Fig.~\ref{Fig1}, the interacting region is a puddle as the one considered in
Refs.~\onlinecite{crep,var,var1}, the effect of the flipping parameter would introduce a significant resistive behavior with the consequent reduction of the conductance in a 6-terminal measurement like the one of Ref.~\onlinecite{Roth:2009bg}. This is an interesting outcome, since such a resistive behavior has been already experimentally observed.

\section{Acknowledgements}
We acknowledge support by CONICET, MINCyT and UBACyT from Argentina. LA thanks the hospitality of ICTP-Trieste and the support of a Simons associateship.

\appendix 
\section{Lesser Green function of the edge in contact to the quantum dot}
\label{sec:appendix}
We follow a similar approach to the one introduced in Ref.~\onlinecite{nos}. We summarize the main
steps to derive the expression for the Green functions of the quantum dot. We  start by defining the following Green functions in the Keldysh contour 
\begin{eqnarray}
i {G}^{\cal C}_{\alpha \sigma, \alpha^{\prime} \sigma^{\prime}}(x,x^{\prime}; t,t^{\prime})  &=& \langle T_{\cal C} [ \Psi_{\alpha,\sigma}(x,t) \Psi_{\alpha^{\prime},\sigma^{\prime}}^{\dagger}(x^{\prime}, t^{\prime}) ] \rangle, \nonumber \\
i { G}^{\cal C}_{\alpha \sigma,  \sigma^{\prime}}(x; t,t^{\prime})  &=& \langle T_{\cal C} [ \Psi_{\alpha,\sigma}(x,t) d_{\sigma^{\prime}}^{\dagger} (t^{\prime}) ] \rangle, \nonumber \\
i {G}^{\cal C}_{\sigma,  \sigma^{\prime}}(t,t^{\prime})  &=& \langle T_{\cal C} [ d_{\sigma}(t) d_{\sigma^{\prime}}^{\dagger} (t^{\prime}) ] \rangle,
\end{eqnarray}
where $T_{\cal C}$ denotes temporal ordering along the Keldysh contour.  These Green functions can be expressed in terms of retarded $G^{R}(t,t^{\prime})= -i \theta (t-t^{\prime}) \langle [ O(t),O^{\dagger}(t^{\prime}) ]_ + \rangle$ and  lesser components 
$G^{<}(t,t^{\prime})= i  \langle O^{\dagger}(t^{\prime}) O(t) \rangle$.\cite{jauho}

The contour-ordered functions obey the following Schwinger-Dyson equations, which when Fourier-transformed 
with respect to $t-t^{\prime}$ lead to the following equations
\begin{eqnarray}
{G}^{\cal C}_{\alpha \sigma, \alpha^{\prime} \sigma^{\prime}}(x,x^{\prime}; \omega)  &=&\delta_{\sigma, \sigma^{\prime}} g^{\cal C}_{\alpha\sigma}(x,x^{\prime};\omega)+
{G}^{\cal C}_{\alpha \sigma, \sigma^{\prime}}(x ; \omega)   \nonumber \\
& & \times \frac{v_F \hbar}{\sqrt{d}}\gamma g^{\cal C}_{\alpha^{\prime} \sigma^{\prime}}(x_0,x^{\prime};\omega), \nonumber \\
{G}^{\cal C}_{\alpha \sigma, \sigma^{\prime}}(x ; \omega)  & = &g^{\cal C}_{\alpha \sigma}(x, x_0;\omega) \frac{v_F \hbar}{\sqrt{d}}\gamma {G}^{\cal C}_{\sigma, \sigma^{\prime}}(\omega).
\end{eqnarray}

Substituting the second of these equations into the first one and evaluating the lesser component by
applying the Langreth rules for the complex contour, \cite{jauho} we find the expression for the lesser Green function for the edge channels
\ba
G_{\alpha\sigma,\alpha\sigma}^{<}(x,x^{\prime};\omega)&=&g^{<}_{\alpha\sigma}(x,x^{\prime};\omega)+\frac{(\hbar v_F)^2}{d} \gamma^2 
\left[
g^{}_{\alpha\sigma}(x,x_0;\omega)\right.\nonumber\\
&&\times  \left.G_{\sigma,\sigma}^{R}(\omega)g^{<}_{\alpha\sigma}(x_0,x^{\prime};\omega)+\right.\nonumber\\
&& \left. g^{}_{\alpha\sigma}(x,x_0;\omega)G_{\sigma,\sigma}^{<}(\omega)g^{*}_{\alpha\sigma}(x_0,x^{\prime};\omega)+ \right. \nonumber\\ 
&& \left. g^{<}_{\alpha\sigma}(x,x_0;\omega)G_{\sigma,\sigma}^{A}(\omega)g^{*}_{\alpha\sigma}(x_0,x^{\prime};\omega)\right],\nonumber\\ 
\ea
where $G_{\sigma,\sigma}^{A}(\omega)=G_{\sigma,\sigma}^{R^{*}}(\omega)$ is the Fourier transform of the retarded Green function for the interacting dot coupled to the edge states and $g_{\alpha\sigma}(x, x^{\prime};\omega)$
are the retarded Green functions of the free HES. Explicit expressions for the latter are given in Appendix \ref{sec:appendix0}. $g_{\alpha\sigma}^{<}(x_0,x_0;\omega)=-if_{\alpha\sigma}(\omega)\,2\,\mathcal{I}m \lbrace g_{\alpha\sigma}^{}(x_0,x_0;\omega)\rbrace$ is the lesser Green function of the isolated HES and $f_{\alpha\sigma}(\omega)$ is the Fermi distribution that defines the filling of each edge channel $\alpha,\sigma$. 

In a similar, way, we can derive the following Green function for the dot in the Keldysh contour
\ba
G^{\cal C}_{\sigma,\sigma^{\prime}} (\omega)= g_{\sigma,\sigma^{\prime}}^{\cal C}(\omega) +
 \sum_{\sigma^{\prime \prime}} 
 G^{\cal C}_{\sigma,\sigma^{\prime \prime}} ( \omega) 
 \Sigma^{\cal C}_{tot, \sigma^{\prime \prime}} ( \omega)
 g^{\cal C}_{\sigma^{\prime \prime},\sigma^{\prime}} ( \omega),\nonumber
\ea
where the total self-energy is $\Sigma^{\cal C}_{tot, \sigma} ( \omega) =\Sigma^{\cal C}_{0, \sigma} ( \omega)+ \Sigma^{\cal C}_{\sigma} ( \omega)$. The term 
\be
\Sigma_{0, \sigma}^{\cal C}(\omega)=  \left(\frac{\hbar v_F}{\sqrt{d}}\gamma\right)^2  \sum_{\alpha=R,L}g_{\alpha\sigma}^{\cal C}(x_0,x_0;\omega),
\ee
is due to the coupling to the edge states and the term $\Sigma^{\cal C}_{\sigma} ( \omega)$  is due to the interaction $U$. 
By recourse to Langreth rules, we can calculate the lesser component of this function
\be
G^{<}_{\sigma ,\sigma^{\prime}} (\omega)= 
 \sum_{\sigma^{\prime \prime}} 
 G^{R}_{\sigma,\sigma^{\prime \prime}} ( \omega) \left[ \Sigma^{<}_{0, \sigma^{\prime \prime}} ( \omega) +
 \Sigma^{<}_{\sigma^{\prime \prime}} ( \omega) \right]
 G^{A}_{\sigma^{\prime \prime},\sigma^{\prime}} ( \omega),
\ee
with $G^{A}_{\sigma^{\prime \prime}, \sigma^{\prime}} ( \omega)= \left[ G^{R}_{\sigma^{\prime}, \sigma^{\prime \prime}} ( \omega) \right]^*$.

\section{Retarded Green function of the isolated helical edge states}
\label{sec:appendix0}

The Green functions of the free HES are given by  
\ba \label{g0r0}
g_{\alpha\sigma}(x, x^{\prime};\omega) &=& 
 \int_{-k_0}^{+k_0}\, dk \, g_{\alpha\sigma}(k;\omega)\, e^{i k(x- x^{\prime})}, \nonumber \\
 g_{\alpha\sigma}(k;\omega) & = & \frac{1}{\omega - v_{\alpha}\hbar k + i\eta}, 
\ea
where $v_{\alpha}\equiv s_{\alpha} v_F$ and $s_{\alpha}=1\,(-1)$ for $\alpha=R\,(L)$. Taking the limit $k_0 \rightarrow \infty$, we obtain the following result:
\ba \label{g0r1}
g_{\alpha\sigma}(x, x^{\prime};\omega) &=& \frac{-i}{ v_F \hbar} \Theta\left( s_{\alpha} (x-x^{\prime}) \right) 
e^{i \frac{s_{\alpha}}{v_F\hbar}\omega  (x- x^{\prime})}.
\ea

\end{document}